\documentclass[twocolumn,twoside,superscriptaddress,prl,aps]{revtex4}
\usepackage{amsmath,mathrsfs,amsbsy,color,graphicx,bm,amsthm,amsfonts}
\usepackage{units}
\usepackage{bbm}

\begin{document}
\title{Tunable single-photon frequency conversion in a Sagnac
interferometer}
\author{Wei-Bin Yan}
\affiliation{Beijing National Laboratory for Condensed Matter
Physics, Institute of Physics, Chinese Academy of Sciences, Beijing
100190, China}
\author{Jin-Feng Huang}
\affiliation{State Key Laboratory of Theoretical
Physics, Institute of Theoretical Physics, Chinese Academy of
Sciences, Beijing 100190, China}
\author{Heng Fan}
\email{hfan@iphy.ac.cn}
\affiliation{Beijing National Laboratory for Condensed Matter
Physics, Institute of Physics, Chinese Academy of Sciences, Beijing
100190, China}
\begin{abstract}
We study the single-photon frequency conversion
of a five-level emitter
coupled to a Sagnac interferometer.
We show that the unity conversion efficiency can be
achieved either in resonance or off-resonance case under the ideal
condition. In particular, the frequency of the output photon can be controlled by the
frequencies and Rabi frequencies of the external driving fields.
\end{abstract}

\pacs{42.50.Ct, 42.65.-k}
\maketitle

Quantum frequency conversion \cite{kumar,kumar1} is a nonlinear process
transducing an input beam of light with a given frequency into an output
beam of light with another different frequency. The quantum frequency
conversion is an indispensable resource for connecting the quantum systems
with different frequencies \cite{wallquist}.
It can be applied for conversion of the photon frequency from visible band to
telecommunication band \cite%
{Mccutcheon,Take,Radnaev,McGuinness,Zaske1,zaske,Ikuta,Greve} or from
telecommunication band to experimentally well detectable band \cite%
{Paulina,Langrock,jomp,ma,matthwe,van,Albota,tan,Clark2013ol}.
The single-photon frequency conversion has many critical applications in
quantum communication and quantum information
processing \cite{gibbs,Bouwmeester}. The highly efficient photon frequency
conversion can be achieved in the large-flux limit \cite{roussev,vandevender}
and at the low-light level \cite{Obiprl2012}. The photon frequency conversion
depends on the nonlinear medium and
the efficient tunable frequency conversion
has not be achieved.

Here we propose a scheme to achieve the efficient controllable single-photon
frequency conversion by adjusting the parameters of the system.
The novel point of our protocol is that for an input photon with a given
wavelength, the wavelength of the converted output photon can be tuned in a
large range. Especially, when the frequency of the output photon is tuned
higher than the input photon, the up conversion is achieved, while the down
conversion can be achieved in the opposite situation. We demonstrate this
control with a five-level emitter coupled to a Sagnac Interferometer\cite%
{Marlan,sagnac1,sagnac2,sagnac3}. The frequency conversion efficiency of the
Sagnac interferometer coupled to a three-level emitter can achieve unity only
when the coupling strengths between the different atomic transitions to the
waveguide loop of the Sagnac interferometer are equal in the resonance case %
\cite{Obiprl2012} for the monochromatic input light, without considering the
dissipation. Here in our system, we show that the frequency conversion efficiency can be
unity in either resonance or off-resonance case. In the resonance case, the
condition under which the different atomic transition-waveguide loop
coupling strengths are equal is not necessarily essential to get a high
conversion efficiency. This is more realizable under practical conditions.
Particularly, the off-resonance case is essential for the control of the
output photon frequency because the frequencies of the input and output
photons are fixed in the resonance case.

\begin{figure}[tbp]
\includegraphics*[width=4.5cm, height=4 cm]{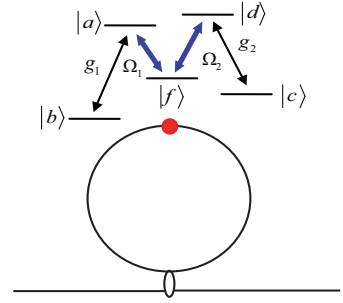}
\caption{A Sagnac interferometer coupled to a five-level emitter. Two
external classical fields are employed to drive the atomic transitions.}
\end{figure}

The structure of the system under consideration is shown in Fig. 1.
The Sagnac interferometer which creates a superposition of two
counter-propagating photon states when a single photon is injected
into the setup consists of a 50:50 coupler and a waveguide loop. To
avoid the output photon returning to the light source, a
supplementary route is necessary as shown in \cite{Obiprl2012}. Here
we do not illustrate this route. The emitter can be a real atom or a
manual atom-like object (The emitter will be mentioned as an atom
below). The two atomic long-live states are denoted by $\left|
b\right\rangle $ and $\left| c\right\rangle $, and the
excited states $\left| a\right\rangle $, $\left| d\right\rangle $, and $%
\left| f\right\rangle $. The atomic level frequencies are represented by $%
\omega _{i}(i=a,b,c,d,f)$. The atomic transitions $\left| a\right\rangle
\leftrightarrow \left| b\right\rangle $ and $\left| d\right\rangle
\leftrightarrow \left| c\right\rangle $ are coupled to the photons in the
waveguide loop with strengths $g_{1}$ and $g_{2}$, respectively. We employ
two external classical fields with frequencies (Rabi frequencies) $\omega
_{L_{1}}(\Omega _{1})$ and $\omega _{L_{2}}(\Omega _{2})$ to drive the
atomic transitions $\left| a\right\rangle \leftrightarrow \left|
f\right\rangle $ and $\left| d\right\rangle \leftrightarrow \left|
f\right\rangle $, respectively. In this paper, we first derive the transport
property of the five-level atom coupled to the waveguide loop and then get
the system output state of the atom coupled to the whole Sagnac
interferometer by the scattering matrix to study the controllable
single-photon frequency conversion. The scattering matrix of the Sagnac
interferometer is $S=S_{c}S_{l}S_{c}$, with $S_{c}=\frac{1}{\sqrt{2}}(%
\begin{array}{cc}
1 & 1 \\
1 & -1%
\end{array}%
)$ representing the beam splitter, and $S_{l}=(%
\begin{array}{cc}
0 & 1 \\
1 & 0%
\end{array}%
)$ representing that the photon goes out from another port different
from the previous input port after a round in the waveguide loop.
The waveguide loop can be treated as a one dimensional waveguide.
Recently, people have done many works in an emitter coupled to a one
dimensional waveguide with many different techniques, such as
solving the time-independent Shr\"{o}dinger equation under some
ansatz in real place \cite{Shen2007pra,Lan}, the
Lehmann-Symanzik-Zimmermann reduction approach in momentum space \cite%
{Shi}, and solving the time-dependent Shr\"{o}dinger equation
\cite{Liao}. By the way, the transverse size effect of waveguide is
first studied in Ref. \cite{Huang}. Here we use the first technique
to get the scattering matrix of the waveguide loop.

The time-independent Hamiltonian of the atom coupled to a waveguide reads ($%
\hbar =1,$ the photonic group velocity $v_{g}=1$)%
\begin{eqnarray}
H &=&\sum_{i}\omega _{i}^{\prime }\sigma ^{ii}-i\int dxa_{e}^{\dagger
}(x)\partial _{x}a_{e}(x) \\
&&-i\int dxa_{o}^{\dagger }(x)\partial _{x}a_{o}(x)+[\sqrt{2}g_{1}\int
dx\delta (x)a_{e}(x)\sigma ^{ab}  \notag \\
&&+\sqrt{2}g_{2}\int dx\delta (x)a_{e}(x)\sigma ^{dc}+\Omega _{1}\sigma
^{af}+\Omega _{2}\sigma ^{df}+h.c.]\text{,}  \notag
\end{eqnarray}%
with $\omega _{f}^{\prime }=\omega _{f}+\omega _{L_{1}},\omega
_{d}^{\prime }=\omega _{d}+(\omega _{L_{1}}-\omega _{L_{2}}),\omega
_{c}^{\prime }=\omega _{c}+(\omega _{L_{1}}-\omega _{L_{2}})$ and
$\sigma ^{ij}=\left| i\right\rangle \left\langle j\right| $ denoting
the atomic raising, lowering and energy level population operators.
It can be seen that the external fields shift the atomic levels. The
expressions of the even and odd operators are
$a_{e}(x)=\frac{1}{\sqrt{2}}[a_{R}^{\dagger }(x)+a_{L}^{\dagger
}(-x)]$ and $a_{e}(x)=\frac{1}{\sqrt{2}}[a_{R}^{\dagger
}(x)-a_{L}^{\dagger }(-x)]$, with the operator $a_{R}^{\dagger }(x)$ and $%
a_{L}^{\dagger }(x)$ creating a clockwise and counterclockwise moving photon
in the waveguide, respectively. Note that the effective atomic frequency $%
\omega _{cb}^{\prime }$ is $\omega _{c}-\omega _{b}+(\omega
_{L_{1}}-\omega _{L_{2}})$ relating to the external field
frequencies. We assume that, initially, the atom is in the state
$\left| b\right\rangle $, and a photon with the wave number $k$ is
injected into the waveguide loop. After scattering, the atom is in
the state $\left| b\right\rangle $ or $\left| c\right\rangle $, with
the corresponding wave number of the output photon $k$ and
$k^{\prime }$, respectively. The former corresponds to the elastic
scattering and the latter to the inelastic scattering. For the
inelastic scattering, the frequency of the output photon depends on
the external field frequencies. Therefore, it is essential to make
sure that the input photon is merely inelastically scattered for
various values of the external field frequencies to achieve the
tunable frequency conversion. The one-excitation state of the
waveguide-atom system can be written as
\begin{eqnarray}
\left| \Psi \right\rangle  &=&[\int dxB(x)a_{e}^{\dagger }(x)+\int
dxC(x)a_{e}^{\dagger }(x)\sigma ^{cb} \\
&&+A\sigma ^{ab}+F\sigma ^{fb}+D\sigma ^{db}]\left| b,0\right\rangle \text{,}
\notag
\end{eqnarray}%
where $B(x)$, $C(x)$,$\ A$,$\ F$, and $D$ are amplitude probabilities, and $%
\left| b,0\right\rangle $ represents that the atom is in the state $\left|
b\right\rangle $ and the photon number in the waveguide is zero. Under the
ansatz $B(x)=[\theta (-x)+t_{1}\theta (x)]e^{ikx}$ and $C(x)=t_{2}\theta
(x)e^{ik^{\prime }x}$, we can get the solution of the time-independent
Shr\"{o}dinger equation $H\left| \Psi \right\rangle =E\left| \Psi
\right\rangle $. The stationary state evolves with time as $\left| \Psi
(t)\right\rangle =e^{-iEt}\left| \Psi \right\rangle $. After calculation %
\cite{zheng2010pra}, the transport properties are obtained as%
\begin{equation}
t_{1}=\frac{A}{B}\text{, }t_{2}=\frac{2i\sqrt{\Gamma _{1}\Gamma _{2}}\Omega
_{1}\Omega _{2}}{B}\text{,}
\end{equation}%
with%
\begin{eqnarray*}
A &=&\Delta _{{a}}(\Delta _{{a}}-\Delta _{1})(\Delta _{{a}}-\Delta
_{1}+\Delta _{2})-\Omega _{{2}}^{2}\Delta _{{a}} \\
&&+\Gamma _{1}\Gamma _{2}(\Delta _{{a}}-\Delta _{1})-\Omega _{{1}%
}^{2}(\Delta _{{a}}-\Delta _{1}+\Delta _{2}) \\
&&+i[\Gamma _{1}(\Delta _{{a}}-\Delta _{1})(\Delta _{{a}}-\Delta _{1}+\Delta
_{2}) \\
&&-\Gamma _{2}(\Delta _{{a}}-\Delta _{1})\Delta _{{a}}-\Omega _{{2}%
}^{2}\Gamma _{1}+\Omega _{{1}}^{2}\Gamma _{2}] \\
B &=&\Delta _{{a}}(\Delta _{{a}}-\Delta _{1})(\Delta _{{a}}-\Delta
_{1}+\Delta _{2})-\Omega _{{2}}^{2}\Delta _{{a}} \\
&&-\Gamma _{1}\Gamma _{2}(\Delta _{{a}}-\Delta _{1})-\Omega _{{1}%
}^{2}(\Delta _{{a}}-\Delta _{1}+\Delta _{2}) \\
&&-i[\Gamma _{1}(\Delta _{{a}}-\Delta _{1})(\Delta _{{a}}-\Delta _{1}+\Delta
_{2}) \\
&&+\Gamma _{2}(\Delta _{{a}}-\Delta _{1})\Delta _{{a}}-\Omega _{{2}%
}^{2}\Gamma _{1}-\Omega _{{1}}^{2}\Gamma _{2}]
\end{eqnarray*}%
where $\Delta _{{a}}=\omega _{a}-k$, $\Delta _{1}=\omega _{af}-\omega
_{L_{1}}$, $\Delta _{2}=\omega _{df}-\omega _{L_{2}}$, and $\Gamma _{m}=%
\frac{g_{m}^{2}}{v_{g}}$ ($m=1,2$) representing the atomic decay rate into
the waveguide loop due to the coupling. Going back to the clockwise and counterclockwise picture
from the even and odd picture, the scattering matrix of the emitter
coupled to the waveguide loop can be derived from $t_{1}$ and $t_{2}$ \cite%
{Shen2007pra} easily and then the whole system scattering matrix can
be calculated. As long as any one of the Rabi frequencies \{$\Omega
_{1},\Omega _{2}\}$ is zero, the frequency conversion efficiency is
zero due to the fact that the atomic transition $\left|
d\right\rangle \leftrightarrow \left| c\right\rangle $ decouples
from the photon in the waveguide and hence the inelastic scattering
vanishes. It is more detailed that when $\Omega _{1}=0$, we can get
$t_{1}={\frac{\Delta _{{a}}+i\Gamma _{1}}{\Delta _{{a}}-i\Gamma
_{2}}}$ and $t_{2}=0$, which is the same as a two-level system
coupled to the waveguide
\cite{Shen2005ol,Shen2007pra,Chang2007nature}. And when $\Omega
_{2}=0$, we can get $t_{1}={\frac{\Delta _{{a}}(\Delta _{{a}}-\Delta _{1})-{%
\Omega _{{1}}}^{2}+i(\Delta _{{a}}-\Delta _{1})\Gamma _{1}}{\Delta _{{a}%
}(\Delta _{{a}}-\Delta _{1})-{\Omega _{{1}}}^{2}-i(\Delta
_{{a}}-\Delta _{1})\Gamma _{1}}}$ and $t_{2}=0$, corresponding to a
three lambda-level atom coupled to the waveguide
\cite{Witthaut2010njp}. This also reveals that the frequency
conversion can be switched off by shutting off the external
classical field, which is equivalent to the control of the relative
phase shift between the clockwise and counterclockwise moving photon
(when the relative phase is $\pi $, an odd-mode quasi particle is
prepared in the waveguide loop and the destructive interference
makes the frequency conversion efficiency zero).

The relationship $\left| t_{1}\right| ^{2}+\left| t_{2}\right|
^{2}=1$ can be easily checked. The maximal frequency conversion
efficiency is $\frac{1}{2}$ when a photon moves only clockwise or
only counterclockwise towards the atom in the waveguide loop. In
this case, the output state has the form of $\left| \Psi
\right\rangle =\frac{1}{\sqrt{2}}\left| b,1_{k}\right\rangle
+\frac{1}{\sqrt{2}}e^{i\phi }\left| c,1_{k^{\prime }}\right\rangle $
($\phi $ is a real number), which is a maximally entangled state.

For an input photon split by the 50:50 coupler, the superposition of
the clockwise and counterclockwise moving states can be prepared in
the waveguide loop. In this case, the interference resulting from
the superposition has a constructive effect on the inelastic
scattering and a destructive effect on the elastic scattering. Once
the relative phase between the photonic clockwise and
counterclockwise moving states is zero,
the scattered state can be obtained as%
\begin{equation}
\left| \Psi \right\rangle =t_{1}\left| b,1_{k}\right\rangle +t_{2}\left|
c,1_{k^{\prime }}\right\rangle \text{.}
\end{equation}%
When $t_{2}=1$, the inelastic scattering process converts the input single
photon into an output photon of the wave number $k^{\prime }$ with unity
conversion efficiency.

\begin{figure}[tbp]
\includegraphics*[width=8.5cm, height=6.5cm]{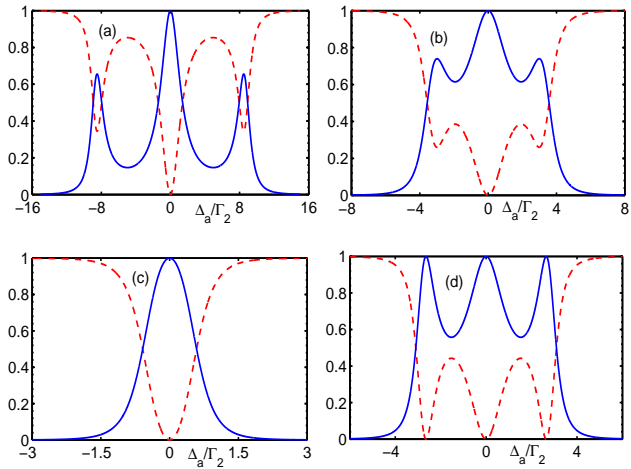}
\caption{Frequency efficiency conversion properties $\left| t_{1}\right|
^{2} $ and $\left| t_{2}\right| ^{2}$ against the input-photon frequency
when $\Gamma _{1}/\Gamma _{2}=\Omega _{1}^{2}/\Omega _{2}^{2}$ and $\Delta
_{1}=\Delta _{2}=0$. The red dashed lines are $\left| t_{1}\right| ^{2}$ and
the blue solid lines are $\left| t_{2}\right| ^{2}$. The parameters are (a)$%
\Gamma _{1}=2\Gamma _{2}$, $\Omega _{1}=5\protect\sqrt{2}\Gamma _{2}$, (b)$%
\Gamma _{1}=2\Gamma _{2}$, $\Omega _{1}=2\protect\sqrt{2}\Gamma _{2}$, (c)$%
\Gamma _{1}=2\Gamma _{2}$, $\Omega _{1}=0.5\protect\sqrt{2}\Gamma _{2}$, (d)$%
\Gamma _{1}=\Gamma _{2}$, $\Omega _{1}=2\Gamma _{2}$. }
\end{figure}

The control of the frequency of the output photon for a high conversion
efficiency is our prime concern. The frequency of the output photon after
the inelastic scattering is obtained as $\omega ^{\prime }=\omega -[\omega
_{cb}+(\omega _{L_{1}}-\omega _{L_{2}})]$ which can be controlled by tuning
the frequencies of the external lasers. This can be understood by the energy
conservation. When $\omega _{cb}+(\omega _{L_{1}}-\omega _{L_{2}})>0$, the
down conversion can be achieved after the inelastic scattering, and when $%
\omega _{cb}+(\omega _{L_{1}}-\omega _{L_{2}})<0$, the up conversion
can be achieved. Obviously, if the resonance condition is satisfied
i.e., $\Delta _{a}=\Delta _{1}=\Delta _{2}=0$, we can get the unity
conversion efficiency when $\frac{\Gamma _{1}}{\Gamma
_{2}}=\frac{\Omega _{1}^{2}}{\Omega _{2}^{2}} $.\ The coupling
strength $g_{1}$ is usually different from the other strength
$g_{2}$ because they depend on the atomic dipole. Hence, the
controllable Rabi frequencies enable us to get a unity conversion
efficiency in the resonance case. Fig. 2 shows the conversion
properties $\left| t_{1}\right| ^{2}$ and $\left| t_{2}\right| ^{2}$
against the frequency of the input single photon when the external
lasers drive the atomic transitions resonantly when $\frac{\Gamma
_{1}}{\Gamma _{2}}=\frac{\Omega _{1}^{2}}{\Omega _{2}^{2}}$. For
small Rabi frequencies, the spectra are shaped like the Lorentzian
line. The spectra split with the increasing Rabi frequencies. When
$\Gamma _{1}=\Gamma _{2}=\Gamma $, and $\Omega
_{1}^{2}=\Omega _{2}^{2}=\Omega $, we can get $t_{1}=\frac{\Delta _{{a}%
}(\Delta _{{a}}^{2}-2\Omega ^{2}+\Gamma ^{2})}{\Delta _{{a}}^{3}-2\Omega
^{2}\Delta _{{a}}-\Gamma ^{2}\Delta _{{a}}-2i(\Gamma \Delta _{{a}%
}^{2}-\Gamma \Omega ^{2})}$. Obviously, when $\Gamma ^{2}-2\Omega ^{2}\geq 0$%
, the unity conversion efficiency can be achieved only when the input photon
interacts with the atom resonantly. However, when $\Gamma ^{2}-2\Omega
^{2}\leq 0$, the unity conversion efficiency can also be obtained even when
the input photon is off-resonant to the atomic transition as shown in Fig.
2(d).

In the case discussed above, the external classical frequencies are
fixed and then can not be tuned to satisfy the resonance condition.
In order to achieve the tunable frequency of the converted output
photon, the unity conversion efficiency in the off-resonance case is
required. In the detuned
case, the condition $t_{1}=0$ requires%
\begin{eqnarray}
\Omega _{1}^{2} &=&\frac{\Gamma _{2}(\Delta _{{a}}-\Delta _{1})({\Delta }%
_{a}^{2}+\Gamma _{1}^{2})}{(\Delta _{{a}}-\Delta _{1}+\Delta _{2})\Gamma
_{1}+{\Delta }_{a}\Gamma _{2}}\text{,} \\
\Omega _{2}^{2} &=&\frac{\Gamma _{1}(\Delta _{{a}}-\Delta _{1})[(\Delta _{{a}%
}-\Delta _{1}+\Delta _{2})^{2}+\Gamma _{2}^{2}]}{(\Delta _{{a}}-\Delta
_{1}+\Delta _{2})\Gamma _{1}+{\Delta }_{a}\Gamma _{2}}\text{.}  \notag
\end{eqnarray}%
\begin{figure}[tbp]
\includegraphics*[width=8.5cm,height=3.2cm]{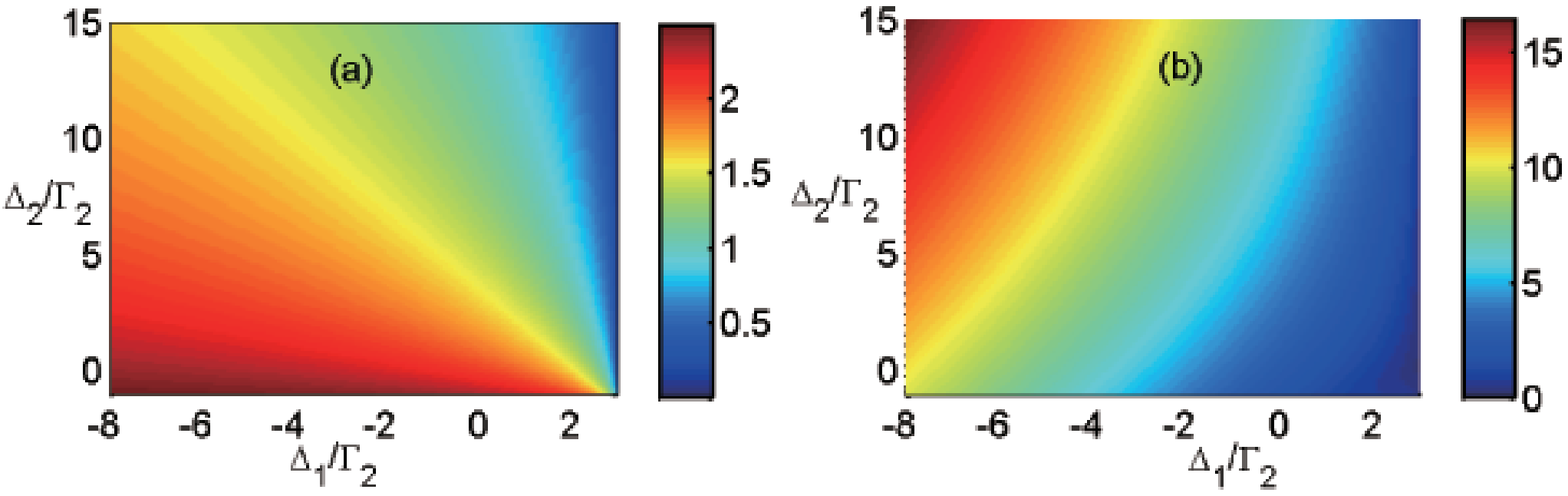}
\includegraphics*[width=8.5cm,height=3.2cm]{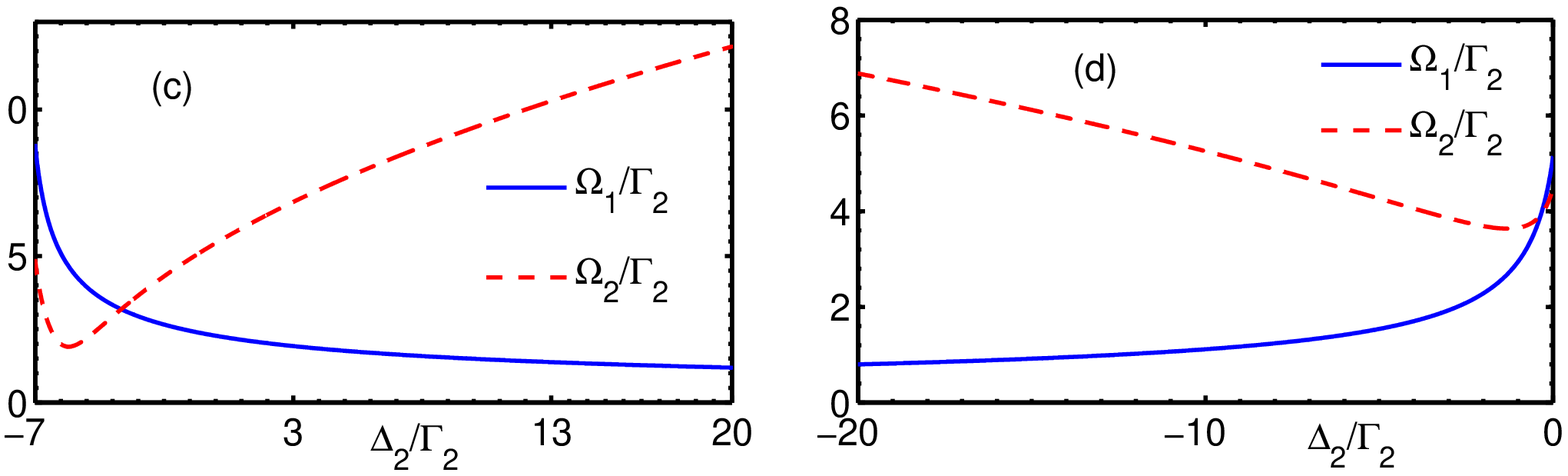}
\caption{The values of Rabi frequencies against the frequencies of external
driving lasers when the photon conversion efficiency is unity. (a) and (b)
are $\Omega_1$ and $\Omega_2$ against the two laser frequencies,
respectively. The parameters are $\Delta _{a}=3\Gamma _{2},\Gamma
_{1}=2\Gamma _{2}$. We take $\Delta _{1}=-3\Gamma _{2}$ in (c), and $\Delta
_{1}=5\Gamma _{2}$ in (d).}
\end{figure}

Therefore, the conditions $\frac{\Gamma _{2}(\Delta _{{a}}-\Delta _{1})({%
\Delta }_{a}^{2}+\Gamma _{1}^{2})}{(\Delta _{{a}}-\Delta _{1}+\Delta
_{2})\Gamma _{1}+{\Delta }_{a}\Gamma _{2}}>0$ and $\frac{\Gamma _{1}(\Delta
_{{a}}-\Delta _{1})[(\Delta _{{a}}-\Delta _{1}+\Delta _{2})^{2}+\Gamma
_{2}^{2}]}{(\Delta _{{a}}-\Delta _{1}+\Delta _{2})\Gamma _{1}+{\Delta }%
_{a}\Gamma _{2}}>0$ are essential to get a unity conversion
efficiency. Although these conditions can not be satisfied for any
arbitrary value of the frequencies of the external fields, they can
be fulfilled in a large range of the frequency values. This feasible
range is enough for the adjusting of the converted-photon frequency
in a wide scale. To explain this, we plot the Rabi frequencies
$\Omega _{1}$ and $\Omega _{2}$ against the frequencies of the
external fields when $t_{1}=0$ in Fig. 3. In Fig. 3(a) and 3(b), we
show the required Rabi frequencies when we adjust both the external
frequencies together. Fig. 3(c) and 3(d)\ show the Rabi frequency
requirement when we adjust one of the external frequency while the
other frequency is fixed. Fig. 3 shows that for the large scale of
the external laser frequencies, the essential conditions above can
be satisfied and the appropriate values of the Rabi frequencies can
be found. Therefore, we can control the frequency of the converted
output photon by controlling the frequencies of the external laser
and tune the Rabi frequencies to get a unity conversion efficiency.
Although the injected photon is not resonant with the atom, the
suitable parameters of the external lasers can ensure the conversion
complete. The frequency conversion process can be understood as a
photon trapping process. After the inelastic
scattering, the injected photon $a$ is trapped and the atom is in the state $%
\left| c\right\rangle $, with another photon $b$ created. Besides, the
trapped photon can be retrieved by injecting the photon $b$.\ It means that,
the photon is trapped for a complete conversion. The retrieval processing
corresponds to the complete conversion $b\rightarrow a$. The retrieval
efficiency can be computed when the atomic initial state is $\left|
c\right\rangle $ by a similar calculation done above. Obviously, the
retrieval efficiency can be unity under a suitable condition.

\begin{figure}[tbp]
\includegraphics*[width=8.5cm, height=6.3cm]{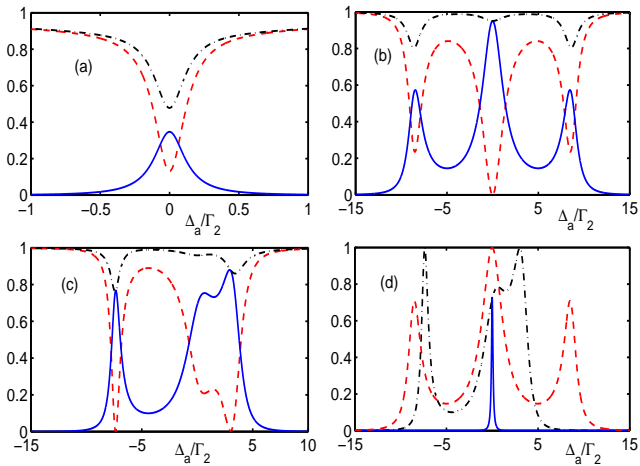}
\caption{The photon conversion properties influenced by dissipation.
(a), (b) and (c) show the probabilities $\left| t_{1}\right| ^{2}$,
$\left| t_{2}\right| ^{2}$, and the total probability $\left|
t_{1}\right| ^{2}+\left| t_{2}\right| ^{2}$, which correspond to the
blue solid lines, red dashed lines, and black dashed dotted lines,
respectively. (d) shows the probability $\frac{\left| t_{2}\right|
^{2}}{\left| t_{1}\right| ^{2}+\left| t_{2}\right| ^{2}}$. The blue
solid line, red dashed line, and black dashed dotted line denote the
situation as shown in (a), (b) and (c), respectively. For all the
plots, the dissipation rate is taken $\protect\gamma =0.1$ and the
coupling strengths $\Gamma _{1}=2\Gamma _{2}$. The respective
parameters are
(a) $\Omega _{1}=\frac{\protect\sqrt{2}}{5}\Gamma _{2}$, $\Omega _{2}=\frac{1%
}{5}\Gamma _{2}$, $\Delta _{1}=\Delta _{2}=0$, (b) $\Omega _{1}=5\protect%
\sqrt{2}\Gamma _{2}$, $\Omega _{2}=5\Gamma _{2}$, $\Delta _{1}=\Delta _{2}=0$%
, (c) $\Omega _{1}=\frac{\protect\sqrt{91}}{3}\Gamma _{2}$, $\Omega _{2}=%
\frac{\protect\sqrt{140}}{3}\Gamma _{2}$, $\Delta _{1}=\Delta _{2}=-4\Gamma
_{2}$.}
\end{figure}
The intrinsic dissipation\ is harmful to achieve the unity conversion
efficiency. This dissipation can be incorporated by introducing the
nonhermitian Hamiltonian $H_{non}=-i\Sigma _{j=a,f,d}\frac{\gamma _{j}}{2}%
\left| j\right\rangle \left\langle j\right| $ in the quantum jump picture,
with $\gamma _{j}$ being the decay rate to other modes except the mode of
the waveguide loop from the level $\left| j\right\rangle $ for a real atom
and being the decay rate plus dephase rate for a manual atom-like object. As
shown above, a complete conversion can be achieved in the resonance and
off-resonance cases under the ideal condition. Fig. 4(a), 4(b) and 4(c) plot
the conversion properties in both the cases after considering the
dissipation. The strong coupling and large detuning can tolerate the
dissipation better. Fig. 4(d) plots the probability $%
\mathscr{F}=\frac{\left| t_{2}\right| ^{2}}{\left| t_{1}\right| ^{2}+\left|
t_{2}\right| ^{2}}$. The high conversion efficiencies
can be obtained in the case as shown in Fig. 4(b) and Fig. 4(c).
The probability $%
\mathscr{F}$ can be nearly unity which means that the input photon
is dissipated and converted, and little elastic scattering exists.
We note another restricting condition that the Rabi frequencies can
not be too small in order to tolerate the dissipation. Fortunately,
after considering this condition, the tunable frequency scale of the
output photon is little affected, which can be understood from Fig.
3.

In summary, we investigate the tunable single-photon frequency
conversion in the proposed system. The inelastic scattering shifts
the frequency of the input photon. The unity frequency conversion
efficiency can be achieved under some conditions. Especially, in the
off-resonance case, the frequency shift can be tuned by adjusting
the external classical fields. Therefore, the output frequency is
tunable. In the dissipation case, the conversion efficiency can not
achieve unity and the output photon is mostly the inelastically
scattered photon.

This work is supported by ``973" program (2010CB922904), grants from Chinese
Academy of Sciences, NSFC (11175248).

\bigskip
\end{document}